\documentclass[a4paper,11pt]{article}
\pdfoutput=1 

\usepackage{jcappub} 
\usepackage{xcolor}
\usepackage[noabbrev]{cleveref}
\usepackage{xspace}
\usepackage{verbatim}
\usepackage{comment}
\usepackage{amsmath}
\usepackage[normalem]{ulem}

\newcommand{\lya}{Ly$\alpha$\xspace}
\newcommand{\lyaf}{Ly$\alpha$ forest\xspace}

\newcommand{\pcross}{\ensuremath{P_\times}\xspace}
\newcommand{\poned}{\ensuremath{P_{\rm 1D}}\xspace}
\newcommand{\xioned}{\ensuremath{\xi_{\rm 1D}}\xspace}

\newcommand{\pthreed}{\ensuremath{P_{\rm 3D}}\xspace}
\newcommand{\kms}{\ensuremath{{\rm km~s^{-1}}\xspace}}
\newcommand{\ikms}{\ensuremath{{\rm s~km^{-1}}\xspace}}
\newcommand{\Mpc}{\ensuremath{{\rm Mpc}\xspace}}
\newcommand{\iMpc}{\ensuremath{{\rm Mpc^{-1}}\xspace}}

\newcommand{\pfft}{\texttt{picca-fft\,}}

\title{Weighted FFT estimators for 1D and 3D correlations of the Lyman-$\alpha$ forest}

\author[a]{Martine Lokken}
\affiliation[a]{Institut de F\'{\i}sica d'Altes Energies (IFAE), The Barcelona Institute of Science and Technology, 08193 Bellaterra (Barcelona), Spain}
\emailAdd{mlokken@ifae.es}

\author[a]{, Andreu Font-Ribera}
\emailAdd{afont@ifae.es}

\author[b]{, Patrick McDonald}\affiliation[b]{Lawrence Berkeley
National Laboratory, 1 Cyclotron Road, Berkeley, CA 94720,
USA}

\abstract{
  Correlations in the Lyman-$\alpha$ (\lya) forest, both as a function of line of sight separation (1D) and 3D separation, provide a unique window to the distribution of matter at redshifts not accessible by current galaxy surveys. 
  While optimal quadratic estimators have been used to measure 1D correlations, they are computationally expensive and difficult to extend to 3D analyses.
  On the other hand, estimators based on the Fast Fourier Transform (FFT) are significantly faster, but are affected by missing data in the spectra (masked pixels) and so far have not used pixel weights to reduce the uncertainties in the measurement. In this publication we describe how to compute the window matrix that enables forward-modelling the impact of masked pixels and weights on the FFT-based estimators.
  We use Gaussian and hydrodynamical simulations with artificially masked pixels to validate the method on the measurement of 1D correlations.   
  Finally, we show that the formalism can be extended to model the impact on 3D correlations, in particular on the cross-spectrum, the correlation of 1D Fourier modes as a function of transverse separation.
  This work will enable more precise clustering measurements with the \lya forest dataset recently collected by the Dark Energy Spectroscopic Instrument (DESI).
}

\begin{document}
\maketitle

\section{Introduction}
\label{sec:intro}

The three-dimensional power spectrum of the \lya forest contains ample information about the 
distribution of matter in the high-redshift universe, enabling state-of-the-art constraints on the sum of the neutrino masses, the properties of dark matter particles, and on the physics of the intergalactic medium (IGM).
Observations of \lya from redshifts $z\sim2-4$ are improving rapidly as DESI spectroscopically surveys a large sample of quasars, providing the opportunity to test our physical models at an early era of structure formation that is not accessible through galaxy surveys.

However, until recently, the \lya spectra were sparsely distributed on the sky, leading the field to focus on measuring only the large-scale baryon acoustic oscillations in 3D \citep[e.g.,][]{Slosar2011, Busca2013, FontRibera2014, dMdB2020, DESI2024.IV.KP6, DESI_DR2_BAO2025}. 
Small scales have been probed instead by measuring the 1D correlation of spectral pixels within a given quasar spectrum,
called \poned \citep{Croft1998, McDonald2000, McDonald2006, PalanqueDelabrouille2013, Chabanier2019, Ravoux2023, Karacayli2024}. 
Recent expansions in data have also enabled significant measurements of the cross-spectrum \pcross \citep{AbdulKarim2024}, which correlates spectra separated by a given transverse distance, incorporating both small-scale line-of-sight information and a range of small-to-intermediate transverse scales. 
At the time of publication, the Dark Energy Spectroscopic Instrument (DESI) \citep{Snowmass2013.Levi,DESI2016a.Science} has already observed four times more \lya quasars than previous surveys \citep{DESI_DR2_BAO2025}. Given the unprecedented precision possible for \poned and \pcross measurements from this new data, accurate systematics control and modeling of these statistics is essential to obtain unbiased results on cosmological and 
IGM parameters. 


The quasar spectra are affected in many ways before summary statistics are measured. 
Among these, particular wavelengths in the spectra are contaminated by nearby absorption (from Galactic gas and the atmosphere) and emission (from the atmosphere). Once the light arrives to the instrument, the spectrographs' finite resolution smears the signal, the spectra are read out and stored in discrete pixels, and some pixels end up in an untrustworthy state for a variety of reasons. Before measuring the summary statistic, the spectral-line contaminants and bad pixels are typically removed through masking. The effects of resolution, pixelization, and masking can be described, in combination, in a window matrix which transforms the original correlations to the measured correlations \citep[see, e.g.,][]{Hui2001ApJ...552...15H}.

Robust inference with \poned and \pcross must be able to either remove mask effects from the measurement or include them in the theory model prior to comparing measurement with model. The former approach is achieved well by optimal quadratic estimators, as described in \cite{McDonald2006,FontRibera2018,Karacayli2020}, and used in measurements from the DESI Early Data Release (EDR) \cite{Karacayli2024}. 
These estimators are able to make an unbiased measurement by maintaining the spectra in their native pixel space, down-weighting noisy pixels and removing the contributions from masked pixels. However, the optimal quadratic estimator suffers in speed due to the computationally-expensive algebraic calculations. 

The alternative approach is to use fast Fourier transform (FFT)-based estimators, which perform the power spectra measurements by 
averaging the square of the amplitude of the discrete 1D Fourier modes computed with an FFT. 
The gain in speed is especially relevant for the purposes of measuring \pcross, which scales computationally as a function of the number of quasar pairs ($\mathcal{O}(N_q^2)$, where $N_q$ is the number of quasar spectra) rather than the number of quasars like for \poned ($\mathcal{O}(N_q)$). However, if the FFT of spectra are used directly to calculate \poned or \pcross without deconvolving the window function, any masking or weighting of pixels will bias the measurement with respect to the truth, as we will demonstrate in this work.

FFT-based estimators have been used to calculate \poned with \lya data from the Baryon Oscillation Spectroscopic Survey (BOSS, \cite{PalanqueDelabrouille2013}), from its extension eBOSS \cite{Chabanier2019} and from DESI EDR \cite{Ravoux2023}, and to make the first \pcross measurement on real data from eBOSS \citep{AbdulKarim2024}; we will generally refer to the approach in those works as the \pfft estimators as the code is currently implemented in the \texttt{picca} code package\footnote{\url{https://github.com/igmhub/picca}}. 
Previous analyses using \pfft have not included per-pixel weights, nor the computation of a window matrix to model impacts of masking. In these analyses, the impact of masking was approximated as a multiplicative correction that is calibrated using synthetic data. 
Moreover, in order to minimize the impact of masking in the measurements of \pcross, the implementation in \pfft is limited to pairs of \lya spectra where both uniformly cover a particular redshift bin.

It is possible to improve the signal-to-noise from FFT-based measurements, avoiding data cuts and using pixel-based weights, if one can accurately describe the weights' impact on the summary statistics using a window function.
For robust inference, either the theory model must be then convolved with the window function, or the data measurement deconvolved. 
Window function computations have been used extensively in inference of a variety of cosmological data for decades \citep[e.g.][]{HauserPeebles1973ApJ...185..757H, Hui2001ApJ...552...15H, Hivon2002, Wilson2017, deBelsunce2024}. 
In this work, we present the window calculations specifically for FFT estimators of \lya \poned and \pcross. 
We describe estimators that can use all the data, include per-pixel weights, and are configured to enable rapid modeling of the window function.

In \cref{sec:p1d_math}, we present the FFT estimators and the impacts on them from weights and masked pixels. We present the model and demonstrate its accuracy for \poned using Gaussian simulations of 1D spectra. In \cref{sec:periodic_boxes}, we further validate the model for \poned by applying various forms of mask to mock spectra from hydrodynamic simulations. In \cref{sec:px}, we discuss the application to estimators of \pcross and demonstrate a simple example with Gaussian simulations. We conclude in \cref{sec:conclusion}.


\section{Masking in FFT estimators of \poned}
\label{sec:p1d_math}

We start in \cref{ss:p1d_fft} by introducing our notation, and reviewing the FFT algorithm to measure \poned in the absence of pixel weights or masking. 
In \cref{ss:nonperiodic} we use simple Gaussian random fields to show the impact of assuming periodic boundary conditions in the FFT algorithm.
In \cref{ss:p1d_mask}, we model the impact of masking in the measured \poned, and we use the same framework to describe the impact of working with non-periodic \lya skewers.

\subsection{FFT measurements without pixel weights}
\label{ss:p1d_fft}

In order to properly model the measured \poned, we need to understand the relationship between the continuous field and the discrete, pixelized measurements. This section describes that relationship, for the case of FFT estimators without pixel weights or masking.
Let us start with an infinitely long, continuous field $\delta(x)$ describing the fluctuations in the \lya forest along a line-of-sight (we will refer to this as a \lya \textit{skewer}).
Assuming translational invariance (ignoring redshift evolution), we can introduce the one-dimensional (1D) correlation function as the covariance of two fluctuations separated by a distance $r$:
\begin{equation}
    \xioned(r) = \left< \delta(x) ~ \delta(x+r) \right> ~. 
\end{equation}

Let us now define the (one-dimensional) continuous Fourier Transform (FT) of the \lya skewer and its inverse as:
\begin{equation}
 \delta(k) = \int dx ~ e^{-i kx} ~ \delta(x) 
 \qquad , \qquad
 \delta(x) = \int \frac{dk}{2\pi} ~ e^{i kx} ~ \delta(k) ~.
\end{equation}
Using these definitions, the covariance of two Fourier modes can be written as
\begin{equation} \label{eq:p1d_def}
    \left< \delta(k) ~ \delta^\ast(k^\prime) \right> = 2 \pi ~ \delta^D(k-k^\prime) ~ \poned(k) ~, 
\end{equation}
where $\delta^D(k)$ is the Dirac delta function and $\poned(k)$ is the one-dimensional power spectrum, the FT of $\xioned(r)$.

\subsubsection{Discretizing the skewers}

Let us now assume that all the skewers in our survey have the same length $L$ and the same number of pixels $N$, centered around positions $x_a = a \Delta x$ with $\Delta x = L / N$.
We can define the pixelised fluctuations $\delta_a$ as integrals over the continuous fluctuation $\delta(x)$ convolved by a function $R_a(x)$ that takes into account both the pixelisation and the spectral resolution of a given pixel:
\begin{equation}
 \delta_a = \int dx ~ R_a(x-x_a) ~ \delta(x) ~.
\end{equation}
We define $R_a$ with this sign of $x-x_a$, opposite the usual convolution convention, because this is what we expect to have most directly available ---
a quantification of the contribution to each discrete pixel from the continuous true flux, which we naturally store and plot with the contribution to the pixel from shorter wavelengths on the left, longer on the right.

Using the FT definitions above, and the equivalent for $R_a(x)$, we can write
\begin{equation}
 \label{eq:delta_a}
 \delta_a = \int \frac{dk}{2\pi} ~ e^{i kx_a} ~ R_a(-k) ~ \delta(k)  ~,
\end{equation}
and the covariance of two pixels can be described as
\begin{equation}
 \label{eq:cov_ab}
 \left< \delta_a \delta_b \right> = \int \frac{dk}{2\pi} e^{i k (x_a - x_b)} R_a(-k) R_b(k) \poned(k) ~,
\end{equation}
where we expanded $\delta_b$ using the complex conjugate of the integrand of \cref{eq:delta_a} and used $(R_b(k))^*=R_b(-k)$; both are possible because $\delta_b$ and $R_b$ are real.

If the functions $R_a(x)$ are symmetric, then their Fourier transforms are real, and we have $R_a(-k)=R_a(k)$.
Moreover, if we also assume that all pixels have the same resolution, we can use a single function $R_a(x)=R(x)$ for all pixels. 
In this case, \cref{eq:cov_ab} becomes a function of the pixel separation $r_{ab}=|x_b-x_a|$, and it can be expressed as a (smoothed) correlation function: $\langle \delta_a \delta_b \rangle = \xioned^R(r_{ab})$. 
The spectroscopic pipeline of DESI \cite{Spectro.Pipeline.Guy.2023} provides a \textit{resolution matrix} that is different for each pixel, and not necessarily symmetric. For simplicity in this article, we will ignore this and use a single symmetric $R(x)$ function. 

\subsubsection{Discretizing the Fourier transforms}

The next step towards deriving the FFT estimator is to define the Discrete Fourier Transform (DFT) of the pixelized skewers and its inverse as:
\begin{equation} \label{eq:DFT_skewer}
 \delta_m = \sum_a \delta_a ~ e^{-i k_m x_a} ~ 
  \qquad , \qquad
 \delta_a = \frac{1}{N} \sum_m \delta_m ~ e^{ i k_m x_a} ~, 
\end{equation}
with $k_m = m \Delta k$, $\Delta k = 2 \pi / L$ and $m \in [0,N).$
While the $N$ values of $\delta_a$ are real, the $N$ discrete Fourier modes $\delta_m$ are complex. However, not all of them are independent, and they satisfy $\delta_{N-m} = \delta^\ast_m$ 
\footnote{In order to speed up the computation and save the memory footprint we could use \texttt{rfft} function of the \texttt{numpy.fft} package and work with $N/2+1$ independent modes. However, for simplicity in this article we use the \texttt{fft} function, the standard \texttt{numpy} implementation of the Fast Fourier Transform (FFT) algorithm. Note that this returns the modes with $m \in [-N/2,N/2)$, sorted beginning with $m=0$ followed by the positive modes, then the negative ones in ascending order, but we can use $\delta_{-m} = \delta_{N-m}$.}.
The covariance of two of the discrete Fourier modes is:
\begin{equation} \label{eq:delta_m_delta_n}
  \left< \delta_m ~ \delta_n^\ast \right> = \sum_a e^{-i k_m x_a} \sum_b e^{i k_n x_b} ~ \left< \delta_a \delta_b \right> ~,
\end{equation}
where $a , b \in [0,N)$.
At this point we could use $\left< \left| \delta_m \right|^2 \right>$ to define a FFT estimator for \poned. However, \cref{eq:delta_m_delta_n} shows that we would need to do $N^2$ operations in order to compare a given model to the measured power.

\subsubsection{FFT estimators in periodic skewers}

The situation is considerably simpler when dealing with periodic \lya skewers, for instance when working with skewers extracted from hydrodynamic simulations with periodic boundary conditions.
This means that we can extend them outside of the original range by defining $\delta_{a+N}=\delta_a$.
In this case, the correlation function is also periodic, and because it is also symmetric around zero separation, we have that $\xioned^R(r_{ab}) = \xioned^R(L-r_{ab})$.
Even though the \lya skewers in real observations are not periodic, this is a useful exercise that will allow us to introduce some of the relevant concepts used later on.
In \cref{ss:nonperiodic} we will show that assuming periodic conditions on real observations has important consequences that need to be taken into account when comparing measurements and model predictions.

With the assumption of periodic skewers, we can define $c=b-a$, $d=c+N$, and rewrite \cref{eq:delta_m_delta_n} as:

\begin{align} \label{eq:delta_m_delta_n_v2}
 \left< \delta_m ~ \delta_n^\ast \right> 
    = & \sum^N_{a=0} e^{-i k_m x_a} \sum^{N-a}_{c=-a} e^{i k_n (x_a + x_c)} 
            ~ \left< \delta_a \delta_{a+c} \right>       \nonumber \\
    = & \sum^N_{a=0} e^{-i k_m x_a} \bigg( \sum^{N-a}_{c=0} e^{i k_n (x_a + x_c)} 
            ~ \left< \delta_a \delta_{a+c} \right>       
        + \sum^0_{c=-a} e^{i k_n (x_a + x_c)} 
            ~ \left< \delta_a \delta_{a+c} \right> \bigg)      \nonumber \\
    = & \sum^N_{a=0} e^{-i k_m x_a} \bigg( \sum^{N-a}_{c=0} e^{i k_n (x_a + x_c)} 
            ~ \left< \delta_a \delta_{a+c} \right> 
        + \sum^N_{d=N-a} e^{i k_n (x_a + x_d - L)} 
            ~ \left< \delta_a \delta_{a+d-N} \right>  \bigg)     \nonumber \\
    = & \sum^N_{a=0} e^{-i k_m x_a} \bigg( \sum^{N-a}_{c=0} e^{i k_n (x_a + x_c)} 
            ~ \left< \delta_a \delta_{a+c} \right> 
        +  \sum^N_{d=N-a} e^{i k_n (x_a + x_d)} 
            ~ \left< \delta_a \delta_{a+d} \right>    \bigg)   \nonumber \\
    = & \sum^N_{a=0} e^{-i k_m x_a} \sum^N_{c=0} e^{i k_n (x_a + x_c)} 
            ~ \left< \delta_a \delta_{a+c} \right>       \nonumber \\
    = & \sum^N_{a=0} e^{-i (k_m-k_n) x_a} \sum^N_{c=0} e^{i k_n x_c} 
            ~ \xi_{1D}^R(x_c)       \nonumber \\  
    = & N \delta^K_{mn} P_m ~,            
\end{align}
where we have used the following property of the Kronecker delta function
\begin{equation}
 \delta^K_{mn} = \frac{1}{N} \sum_a e^{i (k_n-k_m) x_a} ~,
\end{equation}
and we have introduced the FFT of the (pixelized) 1D correlation function:
\begin{equation}
 P_m = \sum_{c=0}^N e^{-i k_m x_c} ~ \xioned^R(x_c) ~.
\end{equation}

In the limit of very long skewers, one can discretize the integral in \cref{eq:cov_ab} and show that $P_m$ (dimensionless) and the actual $\poned$ (with units of length) are related by:
\begin{align}
 \label{eq:P_m}
 P_m  &= \sum_c e^{-i k_m x_c} \xioned^R(x_c)   \nonumber \\
    &= \sum_c e^{-i k_m x_c} \int \frac{dk}{2\pi} 
        e^{i k x_c} R^2(k) \poned(k)            \nonumber \\
    &\approx \sum_c e^{-i k_m x_c} \frac{\Delta k}{2\pi} \sum_n 
        e^{i k_n x_c} R^2(k_n) \poned(k_n)            \nonumber \\ 
    &= \frac{\Delta k}{2\pi} \sum_n R^2(k_n) \poned(k_n)
        \sum_c e^{-i (k_m - k_n) x_c}      \nonumber \\  
    &= \frac{1}{\Delta x} R^2(k_m) \poned(k_m)
\end{align}

Therefore, ignoring instrumental noise, the standard FFT algorithm to estimate \poned at discrete Fourier mode $m$ is to average over all $N_q$ skewers the square of the amplitude of their $\delta^q_m$ values:
\begin{equation}
 \label{eq:simple_fft}
 \hat P_{1\mathrm{D},m} \equiv \frac{\Delta x}{N ~ R^2(k_m)} \frac{1}{N_q} \sum_{q} \left| \delta^q_m \right|^2 ~.
\end{equation}
For the case of periodic \lya skewers, we can substitute Eq.~\ref{eq:delta_m_delta_n_v2} to demonstrate that the expected value of the estimator is unbiased:
\begin{equation}
 \left< \hat P_{1\mathrm{D},m} \right> = \frac{\Delta x}{N ~ R^2(k_m)} N P_m = \poned(k_m) ~.
\end{equation}

\subsubsection{FFT estimators in finite, non-periodic skewers}
\label{ss:nonperiodic}

In order to demonstrate that the FFT estimator in \cref{eq:simple_fft} is unbiased, we had to assume that the skewers were periodic, enabling the rearrangement of indices leading to \cref{eq:delta_m_delta_n_v2}. Therefore, the estimator is only unbiased in unrealistic scenarios, such as large numerical simulations with very long skewers and periodic boundary conditions. In realistic scenarios with finite, non-periodic skewers, it is biased, and to calculate the expected value of the estimator we must return to \cref{eq:delta_m_delta_n} for the $\delta_m$ substitution:
\begin{align} \label{eq:nonperiodic_expected_value}
 \left< \hat P_{1\mathrm{D},m} \right> 
    &= \frac{\Delta x}{N ~ R^2(k_m)} 
        \left< \left| \delta_m \right|^2 \right>  \nonumber \\
    &= \frac{\Delta x}{N ~ R^2(k_m)} 
        \sum_a e^{-i k_m x_a} \sum_b e^{i k_m x_b} \left< \delta_a \delta_b \right>  \nonumber \\
    &= \frac{\Delta x}{N ~ R^2(k_m)} 
        \int \frac{dk}{2\pi} \left| R(k) \right|^2 \poned(k) 
        \sum_a e^{i (k-k_m) x_a} \sum_b e^{-i (k-k_m) x_b} ~.
\end{align}

\begin{table}
\centering
\begin{tabular}{c|ccc|cc}
Parameter  &  $P_0$ [\kms]  &  $k_0$ [\ikms]  &  $k_F$ [\ikms]  &  $a_{\rm Si}$  &  $r_{\rm Si}$ [\kms] \\
\hline
Value  &  $10.0$  &  $0.01$  &  $0.1$  &  $0.05$  &  $2271.0$  \\
\end{tabular}
\caption{Parameters describing the input power of the Gaussian skewers (first three columns), and the contamination by Silicon absorption (last two columns).}
\label{tab:true_p1d}
\end{table}

Here we will use simple synthetic skewers to visualize this bias from non-periodicity. We generate a Gaussian random field with the following input power spectrum:
\begin{equation}
 \label{eq:true_p1d}
 \poned(k) = P_0 ~ \frac{1+k/k_0}{1+(k/k_0)^2} ~ e^{-(k/k_F)^2} ~,
\end{equation}
with parameter values described in \cref{tab:true_p1d}. 
The scale dependence is a simple attempt to capture the main features of the real \poned: constant power at large scales, slow rise at intermediate scales, followed by a gradual suppression as one goes to smaller scales, before pressure smoothing erases any structure below the Jeans length. 
As we will see in the next section, the assumption of periodicity will cause a convolution of the true \poned, and therefore any sharp feature will be smoothed.
An example of sharp features in \poned are the oscillations introduced by the contamination by Silicon absorption observed in measurements of \poned \cite{McDonald2006}.
This contamination is often modeled with a multiplicative correction to the \lya power, $\left[ 1 + a_{\rm Si}^2 + 2a_{\rm Si} \cos{(k r_{\rm Si})} \right]$, where $a_{\rm Si}$ sets the amplitude of the oscillations and $r_{\rm Si}$ its frequency.
We add this contamination to the Gaussian skewers, using the parameters described in \cref{tab:true_p1d}.

\begin{figure}
\centering \includegraphics[width=0.8\textwidth]{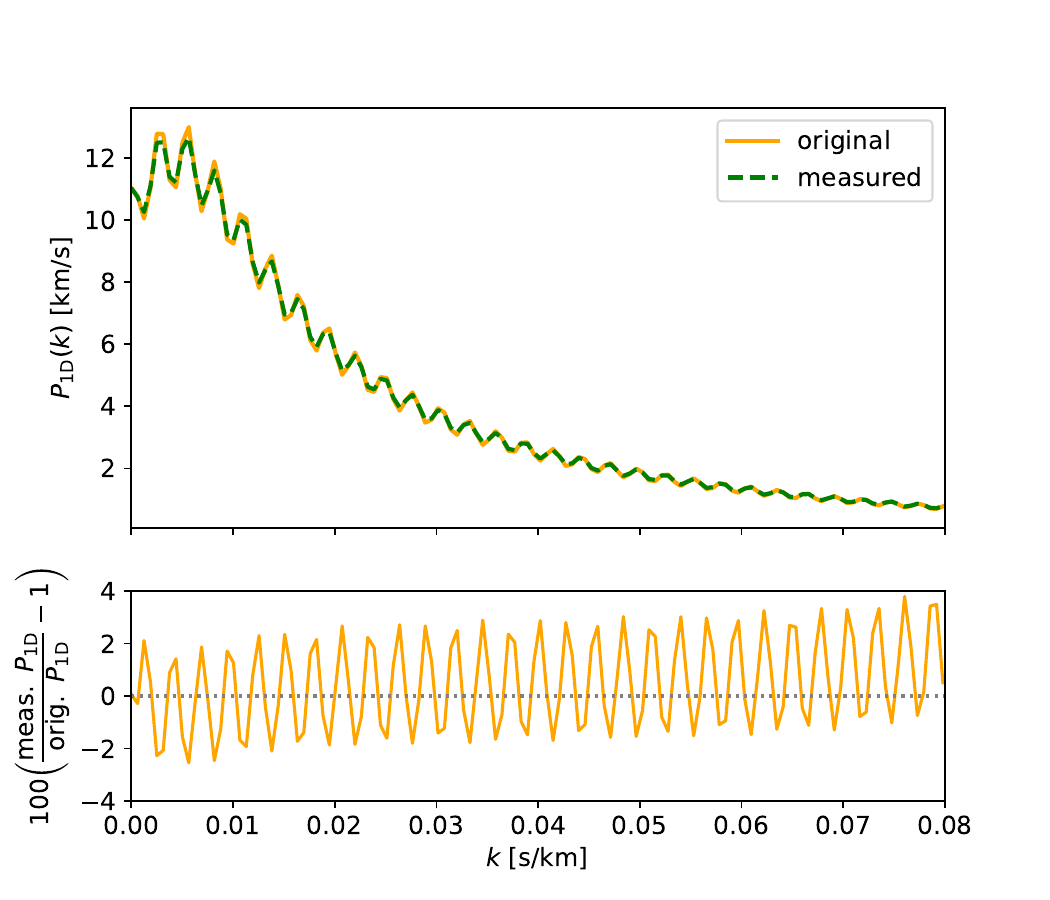}
 \caption{Measurements of \poned in non-periodic Gaussian skewers (green dashed line), compared to the input power (orange).
 The bias is more clearly seen in the bottom panel, which shows the residuals (measurement over original, minus one) as a percentage. The bias appears as both a distortion of the Silicon oscillations and a broad scale dependence increasing with $k$.}
 \label{fig:periodic}
\end{figure}

The latest FFT-based measurement of \poned by the DESI collaboration \cite{Ravoux2023} split each \lya skewer in three \textit{chunks} of approximately $10^4 ~\kms$ in length.
Here we generate a periodic field with $2048$ pixels and a total length of $8 \times 10^4 ~\kms$, but we only keep $1/8$ of the field to obtain a \textit{non-periodic} skewer with $256$ pixels, $10^4 ~ \kms$ long. 
This gives a Fourier mode resolution of $\Delta k = 2\pi/L = 0.0006 ~\kms$.

In \cref{fig:periodic} we show (in green) the measurement of \poned from $10^7$ of these non-periodic skewers, and compare it to the input power used to generate the skewers (in orange).
The residuals shown in the bottom panel demonstrate a clear bias in the measurement of \poned. 
First, the oscillatory features mimicking the contamination by Silicon absorption have been smoothed, causing a bias oscillating by $\sim4\%$. 
Second, the ``broadband" power is slightly under-predicted at low values of $k$ ($k<0.015~\kms$) and over-predicted at intermediate and large values of $k$ (reaching the 1\% level at $k\sim0.07~\kms$).
This second bias is also present in mock spectra without Silicon contamination. 

Conceptually, the two effects arise because the DFT (Eq.~\ref{eq:DFT_skewer}) assumes the signal to be one period of an infinitely repeating signal, and therefore that the final and initial pixel of each spectrum are neighboring. This causes the Fourier signal, and therefore the output power, to become mixed across modes compared to the true input power. The effect is easiest to consider for the correlation function in configuration pixel space: although the initial and final pixels are truly separated by the full length of the spectrum, when the signal is assumed to be periodic the pair contributes to the correlation function bin at a single-pixel length.

One could use \cref{eq:nonperiodic_expected_value} to forward model the impact of non-periodic boundary conditions, but it would be quite tedious computationally. 
It is faster and more convenient to model the impacts after \textit{zero-padding} the signal, as shown in the following section. 
This entails extending the signal by a chain of zeros.
After adding zeros, the estimator is still biased 
\footnote{Imagine extending the signal by zeros such that the new array is twice as long. As evident from Eq.~\ref{eq:DFT_skewer}, besides an overall amplitude suppression from the the null contribution of zeros to a signal with larger $N$, the $\delta_m$ values at the same discrete modes $k_m$ will be exactly equivalent to the values before zero-padding. However, there are now twice as many $k_m$ modes, with the new discrete modes falling at $k$ values between the previous ones. Thus, for extended bins in $k$, even after correcting for the scalar amplitude suppression, the padded signal will be subtly different than the original signal. This is consistent with the fact that $\xi_\mathrm{1D}$ must change.},
but it is easier to forward-model the theoretical predictions for the summary statistics.

Ideally, the zero padding should extend the FFT grid used in all computations to at least twice the original length.
This is because the discretized computations of \poned introduced in this section perform circular convolution of the real-space signal, which introduces a bias at the overlapping edges that have been wrapped around (when applied to a non-periodic signal). Performing each convolution after padding the array with an equal amount of zeros produces $2N-1$ non-zero pixels, and thus avoids overlap between the end and beginning of the array which are separated by $2N$. Longer zero-padding will continue to improve the convergence of the discretized calculations toward the continuous integrals in $k$, but practical limits on memory usage constrain this extension. \poned and \pcross analyses are typically less memory-intensive than full three-dimensional analyses, but in any case, the best practice is to perform tests to determine the minimal zero-padding required to reduce bias from discretization and non-periodicity to a satisfactory level given computational constraints.

\subsection{Modelling the impact of zero-padding, masking, and pixel weights}
\label{ss:p1d_mask}

Zero-padding is only one source of nonphysical values in spectral data; most analyses of the \lya forest also implement some level of masking at the pixel level, motivated by instrumental artifacts or sky lines. Practically, masking is typically done by setting a spectral pixel with an unwanted feature to zero; the mask is thus a specific case of pixel weights, where the weights can only take binary values (ones or zeros). In this section we will model the impact of a general array of pixel weights $w_a$ that can take arbitrary values, with the understanding that a masked pixel will have $w_a=0$.

From now on, we will use $N$ to refer to the total number of pixels in the FFT grid, including the zero padding. We can now define an array of $N$ weighted pixels $f_a = w_a \delta_a$, and compute its FFT:
\begin{equation} \label{eq:f_m}
 f_m = \sum_a e^{-ik_m x_a} ~ w_a ~ \delta_a
    = \int \frac{dk}{2\pi} ~  ~ w(k_m-k) ~ R(k) ~\delta(k)  ~,
\end{equation}
where we have introduced a new function 
\begin{equation}
 w(k) = \sum_a e^{-ikx_a} ~ w_a ~.
\end{equation}
Note that we have not defined $w(k)$ as the continuous Fourier transform of a $w(x)$ function, and that it is a complex number and dimensionless.

The expected value of the (dimensionless) variance of the (discrete) Fourier modes is:
\begin{equation}
 \left< \left| f_m \right|^2 \right> 
    = \int \frac{dk}{2\pi} ~ |w(k_m-k)|^2 ~ R^2(k) ~ \poned(k) ~,
\end{equation}
where the integral reminds us of a convolution of the Fourier transform of the weights with the product of \poned and the resolution kernel.
In 1D analyses it should be easy to add enough zero-padding to make the skewers as long as needed to have a very good frequency resolution ($\Delta k = 2\pi/L$), such that we can discretize this integral with the same set of $k_m$ wavenumbers: 
\begin{equation} \label{eq:fourier_mode_variance}
 \left< \left| f_m \right|^2 \right> 
    = \frac{\Delta k}{2\pi} \sum_n \left| w_{m-n} \right|^2 ~ R^2(k_n) ~ \poned(k_n) 
    = \frac{1}{N} \sum_n \left| w_{m-n} \right|^2 ~ P_n ~,
\end{equation} 
where we have introduced $w_m = w(k_m)$, the DFT of $w_a$, and we have used \cref{eq:P_m} above.

Similarly to \cref{eq:simple_fft}, we can now derive an alternative estimator of \poned to be used in the presence of weights, masking or zero-padding: 
\begin{equation}
 \hat P_{1\mathrm{D},m} \equiv C_m ~ \frac{\Delta x}{N ~ R^2(k_m)} \frac{1}{N_q} \sum_{q}  \left| f^q_m \right|^2  
\end{equation}
(recall $f_m$ is the FT of 
$f_a=w_a \delta_a$ and $N$ is the number of 
pixels in the FFT grid). 
The same normalization factors as in \cref{eq:simple_fft} give dimensions of length\footnote{In \lya analyses, the units of such a `length' could be either km/s, Mpc, or \r{A}.} to \poned, while $C_m$ is an extra normalization factor that we introduce for convenience.

The ensemble average of the estimator is related to the original \poned by:
\begin{align}\label{eq:forward_model}
 \left< \hat P_{1\mathrm{D},m} \right>
    & = C_m ~ \frac{\Delta x}{N ~ R^2(k_m)} \frac{1}{N_q} \sum_{q} \left< \left| f^q_m \right|^2 \right>    \nonumber \\
    & = C_m ~ \frac{\Delta x}{N ~ R^2(k_m)} \frac{1}{N_q} \sum_{q}
        \frac{\Delta k}{2\pi} \sum_n \left| w^q_{m-n} \right|^2 R^2(k_n) \poned(k_n) \nonumber \\
    & = \frac{C_m}{N^2 ~ R^2(k_m)} \sum_n W_{m-n} R^2(k_n) \poned(k_n) \nonumber \\
    & = \sum_n M_{mn} ~ \poned(k_n) ~,
\end{align}
where we have introduced $W_m$ as the average of the squared FFT of the individual weights:
\begin{equation}
 W_m \equiv \frac{1}{N_q} \sum_{q} \left| w^q_m \right|^2 ~,
\end{equation}
and we have introduced the window matrix of the survey:
\begin{equation}
 M_{mn} \equiv \frac{C_m}{N^2 ~ R^2(k_m)} W_{m-n} R^2(k_n) ~.
\end{equation}

The window matrix $M_{mn}$ specifies how much the undistorted model at a given wavenumber $k_n$ contributes to the distorted prediction at a wavenumber $k_m$. 
We can choose to set the normalization factor $C_m$ such that the each row of the window matrix is normalized to 1, i.e., $1 = \sum_o M_{mo}$. 
This condition is satisfied if we choose:
\begin{equation} \label{eq:C_m}
 C_m = \frac{N^2 R^2(k_m)}{\sum_o W_{m-o} R^2(k_o)} ~.
\end{equation}

If we ignore the impact of the resolution kernel (set $R(k)=1$), one can see that for simple masking (binary weights), the normalization factor $C_m$ is simply a scalar corresponding to the inverse of the fraction of pixels not masked. So, $C_m$ can be thought of as a factor rescaling the measured \poned amplitudes (typically lower than the original \poned due to masking) to values similar to the original.

\begin{figure}
\centering
 \includegraphics[width=0.8\textwidth]{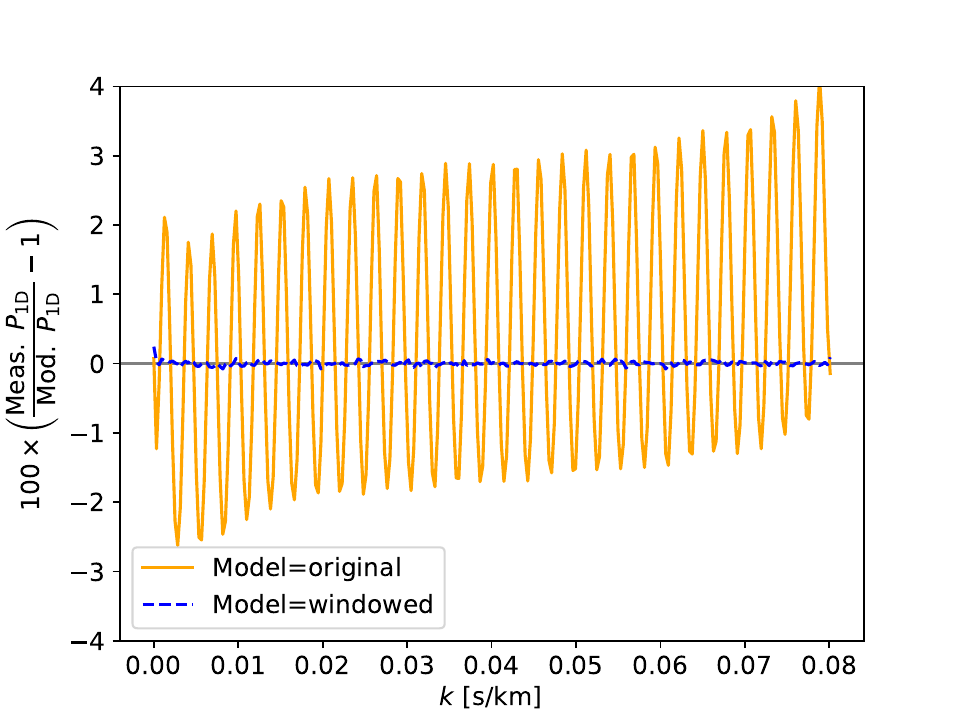}
 \caption{Percentage residuals of the \poned measurement from non-periodic skewers with zero-padding using the estimator in \cref{eq:masked_fft}, compared to the original input \poned (orange) and compared to the model computed using the window matrix in \cref{eq:M_mn} (blue, dashed). Application of the window matrix forward-models the impact of zero-padding on the original input \poned. With this forward modeling, residuals reduce to a negligible level.}
 \label{fig:convolved}
\end{figure}

Using this normalization factor, the estimator can be rewritten as:
\begin{equation}\label{eq:masked_fft}
 \hat P_{1\mathrm{D},m} \equiv \frac{L}{\sum_o W_{m-o} R^2(k_o)} ~ \frac{1}{N_q} \sum_{q}  \left| f^q_m \right|^2 ~,
\end{equation}
and the window matrix as:
\begin{equation}\label{eq:M_mn}
 M_{mn} \equiv \frac{W_{m-n} R^2(k_n)}{\sum_o W_{m-o} R^2(k_o)}  ~.
\end{equation}

Thus, to predict $\left< \hat P_{1\mathrm{D},m} \right>$ from a theory model for the original \poned, one needs to compute the window matrix of the survey only once. The process in practice begins with Fourier transforming the array of real-space weights, then squaring its magnitude and summing over all skewers to get $W_m$. Next, one calculates the window with \cref{eq:M_mn}; the convolution in the denominator can be done rapidly by multiplying the FFTs of $W$ and $R^2$ and transforming back via an inverse FFT. The resulting real-valued, $N\times N$ array is matrix-multiplied by the discretized original \poned theory vector to yield the predicted weighted \poned vector.

We make a measurement of \poned from the same $10^7$ non-periodic skewers used in \cref{fig:periodic}, but this time using the estimator from \cref{eq:masked_fft} and a factor of two zero-padding (we started with the same $256$ pixels, but added now the same amount of pixels with $w_a=0$ such that the FFT grid has $N=512$ pixels). Because the distortion is caused by an input alteration to the signal in a form that we know exactly, we can use \cref{eq:forward_model} to forward-model the distortion. In \cref{fig:convolved} we show the residuals from comparing the measurement to the forward model (blue dashed line) versus comparing the measurement to the original input power (orange). The small residuals using the windowed model demonstrate that the model captures both the smoothing of the oscillatory features and the broadband trend.
 
\section{Validation with masked mock spectra from hydrodynamic simulations} \label{sec:periodic_boxes}

To validate the model on a more realistic set of mock skewers, we move to a suite of \texttt{MP-Gadget} hydrodynamical simulations first presented in \citep{Pedersen2021}. Each simulation box has periodic boundary conditions with a side length $L=67.5~\Mpc$, with $768^3$ particles, and skewers extracted with $N=1350$ pixels along the line of sight, i.e., $\Delta x=0.05~\Mpc$. This pixelisation sets the natural grid to use in the FFTs, and its corresponding wave-numbers are equispaced by $\Delta k=2\pi/L=0.0931 \iMpc$, with a Nyquist frequency of $k_{\rm Ny}=\pi/\Delta x=62.83 \iMpc$. We work with the simulation snapshot at $z=3$.

Because of the periodicity of the hydro boxes, there is no need to use zero-padding; we will apply other forms of realistic masks in this section. As before, we will focus on the effects of these masks on realistic data whose power spectra include localized features in $k$-space. To incorporate such features, we add Silicon contamination using the model introduced in \cref{ss:nonperiodic} and parameters $a_\mathrm{Si}=0.05$, $r_\mathrm{Si}=20$~Mpc. Additionally, we include an arbitrary and simple additional source of contamination, a spike in amplitude localized in Fourier space. 
To do so, we boost the amplitude of a single discrete Fourier mode $\delta_m$ for all skewers,  with $k_m \sim 0.47~\iMpc$, such that the \poned is augmented by 0.5~Mpc for that mode.

We apply five different types of mask to the mock spectra: (a) single emission/absorption line, (b) double emission/absorption line, (c) small DLAs, (d) large DLAs, and (e) random masking to mimic randomly-placed corrupted pixels. In a typical analysis, true \lya data would be masked for a combination of these effects, but we separate them to examine their distinct features. We enforce that each case results in an approximately equal total number of masked pixels: $\sim2\%$ of the total number of pixels summed over all skewers. 

\begin{figure}
    \centering
    \includegraphics[width=0.85\linewidth]{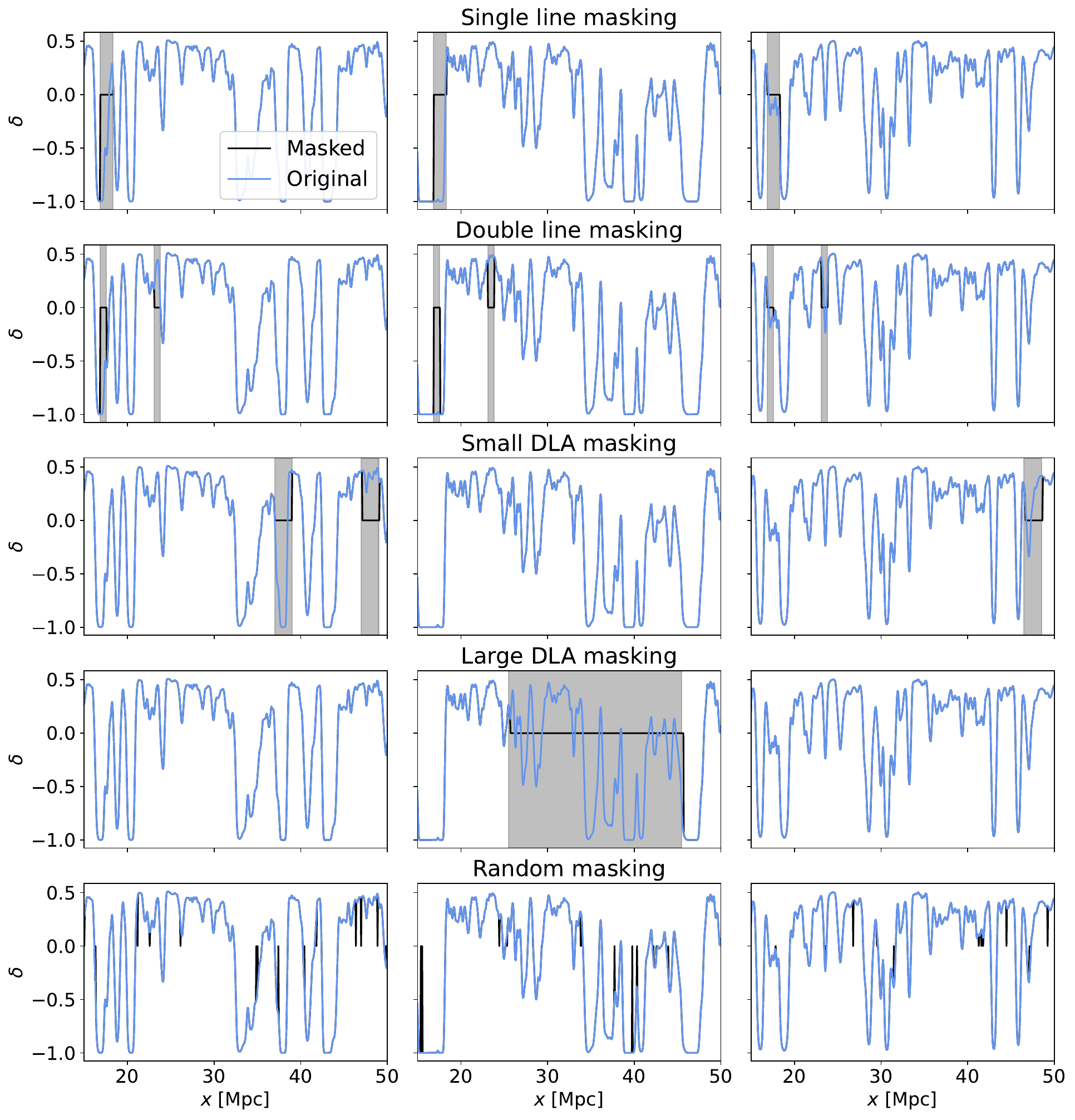}
    \caption{Representation of the five mask schemes (rows) for three mock spectra (columns). The black masked spectra are only visible compared to the over-plotted blue unmasked spectra in the masked regions.  Gray shaded regions indicate the masks, except in the case of random masking where in most cases masked pixels are non-consecutive. The mask always appears in the same location in the mock spectra for single- and double-line masking (upper two rows). For the DLA-like masking, not all spectra host a `DLA', and some host multiple. In all cases, the mask is uncorrelated with the original signal.}
    \label{fig:mask-schema}
\end{figure}

The single-line mask (a) sets thirty consecutive pixels to zero at a particular $x$ value in all skewers. The width of the mask (1.5~Mpc, or $\delta \lambda\sim1.9$\r{A}) is chosen to roughly correspond to the masks for sky emission lines and Galactic Calcium absorption lines used in the DESI \lya early data release 
\citep{Ramirez2024}, as translated into the rest-frame of a $z=3$ quasar. The double-line mask (case b) mimics the single-line mask, but sets 15 consecutive pixels to zero in two locations, separated by 125 pixels (6~Mpc / $\Delta \lambda\sim8$\r{A}).
For DLA-like masking (c) and (d), we choose a subset of 
skewers in which to mask one to two DLA-length sets of consecutive pixels. 
For the large DLA case, we set 20~Mpc (400 consecutive pixels) to zero at a random location along the skewer; for small DLAs, this is reduced to 40 pixels (2~Mpc). We note that the DLA masks are placed randomly, not at absorption peaks. Realistic DLA masking would be correlated with the field, but modeling the impact of correlated masks is beyond the scope of this work; hence we use uncorrelated placement to enable the application of the simple model presented in this paper.
Lastly, in the random mask (e), we mask each skewer by a random number of pixels between 0 and 20 in random locations.  Visualizations of the four masks are shown in \cref{fig:mask-schema}.

\begin{figure}
    \centering
    \includegraphics[width=0.75\linewidth]{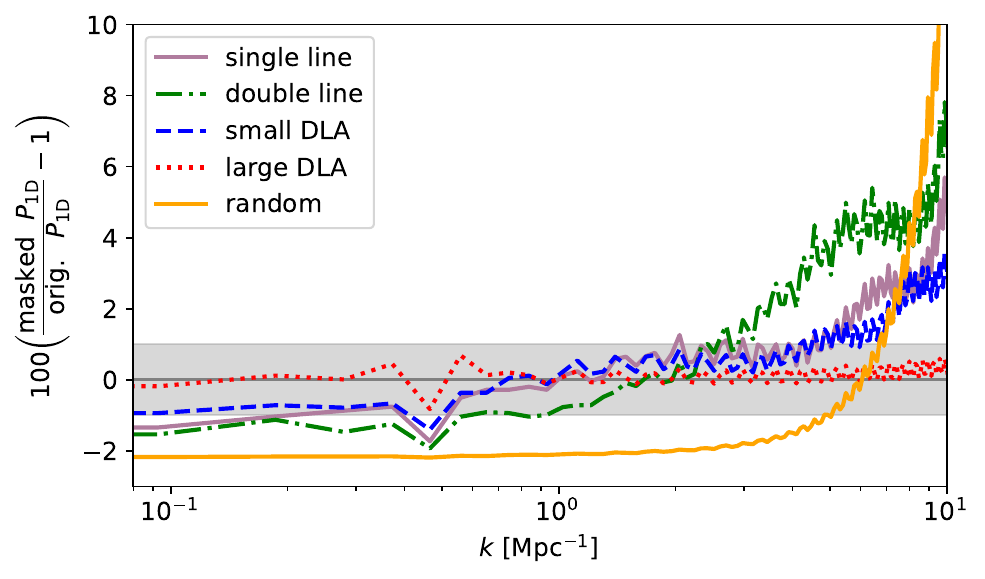}
    \caption{The impacts of the five mask schemes on \poned measured from hydro simulation mock skewers. The \poned estimates are computed by \cref{eq:masked_fft}, and the figure shows the percentage difference between the measurements with masking and without (`orig.').  The `large DLA' case is broadly scale-independent due to the long extent of each DLA mask, but each of the other masking types induce broad $k$-dependent trends which depend on the extent and spacing of the masked pixels. Additionally, in all cases oscillations are clearly visible due to the mask distorting the measurement of the Silicon-contaminated original power spectrum. Distortions around the spike at $k \sim0.5~\iMpc$ are also clear for all but the random mask. Overall, the strongest biases occur at very small scales, but even for a reasonable observational range of $k<3~\iMpc$, all masks except for the large-DLA type cause a bias $>1\%$ (shaded region). In practice, the amount of bias will depend on the amount of masking.}
    \label{fig:mask-impacts}
\end{figure}

We estimate \poned for each set of masked skewers using \cref{eq:masked_fft}, producing a measurement $P_\mathrm{1D,masked}$. We repeat the same estimate with the set of original skewers to produce a measurement without masking effects, $P_\mathrm{1D,orig}$. As both have very similar cosmic variance, their residuals are nearly completely dominated by the mask impacts. \Cref{fig:mask-impacts} shows the residuals. The large DLA masking has a broadly scale-independent impact, but due to their small-scale features, the four other types decrease low-$k$ power and increase high-$k$ power. Moreover, all cases show oscillatory features in $k$ space related to the added Silicon-like contamination. Most also cause a distortion at the location of the spike, again demonstrating that the bias from masking is sensitive to the presence of features in the true power spectrum.

\begin{figure}
    \centering
    \includegraphics[width=0.57\linewidth]{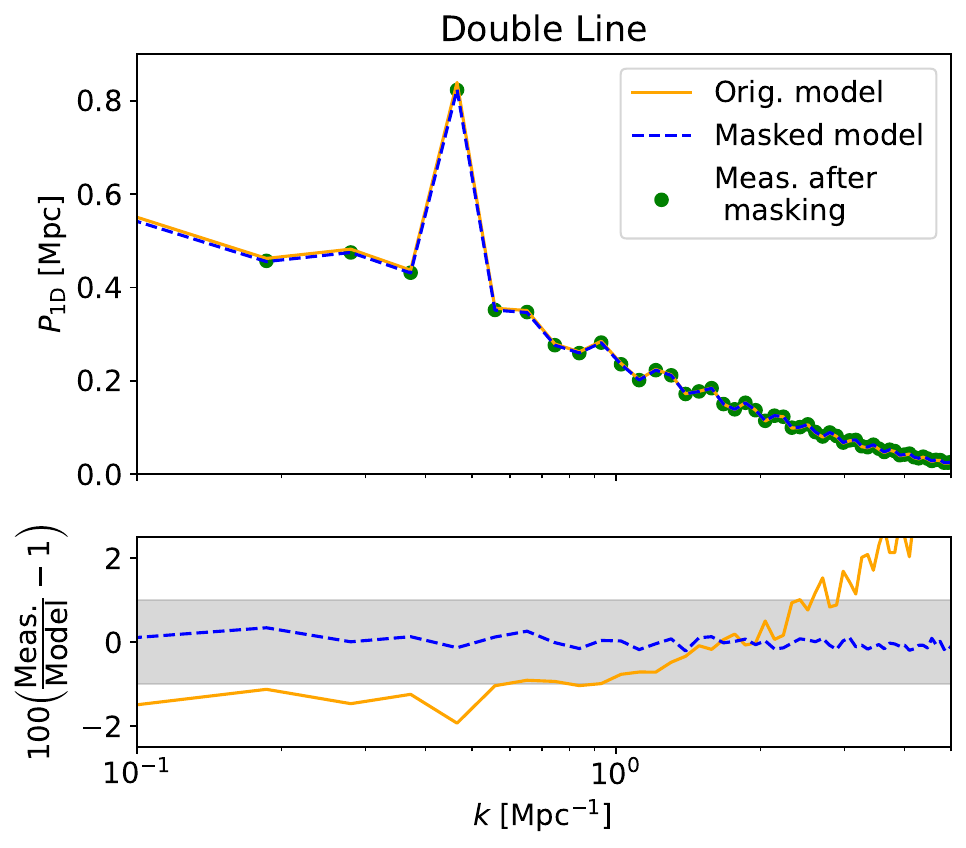} \\
    \includegraphics[width=0.518\linewidth, trim=.25cm 0cm 0.25cm 0.25cm,clip]{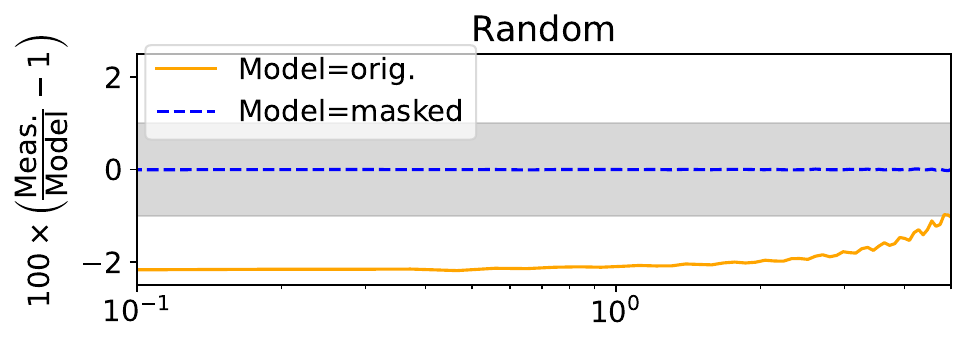}\hspace{0cm} \includegraphics[width=0.473\linewidth, trim=.25cm 0cm 0cm 0.25cm,clip] {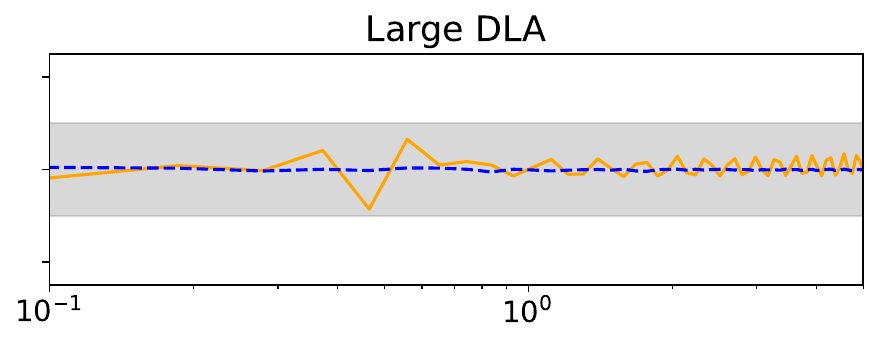}
    \\
    \includegraphics[width=0.518\linewidth, trim=.25cm 0cm 0.25cm 0.25cm,clip]{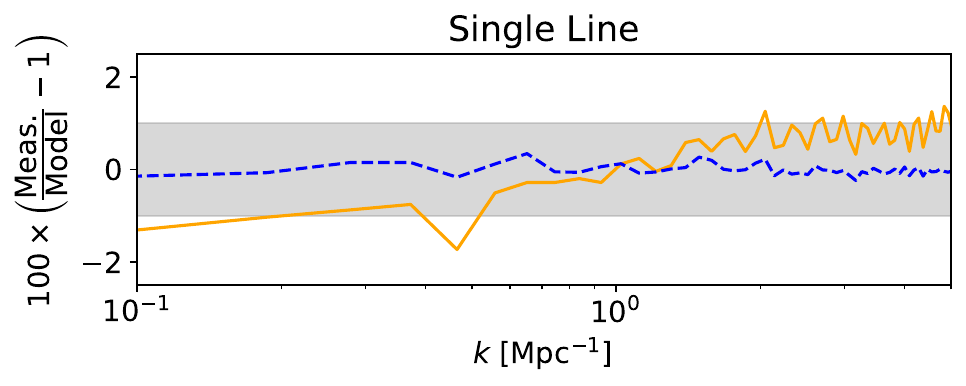}
    \includegraphics[width=0.473\linewidth, trim=.25cm 0cm 0cm 0.25cm,clip] {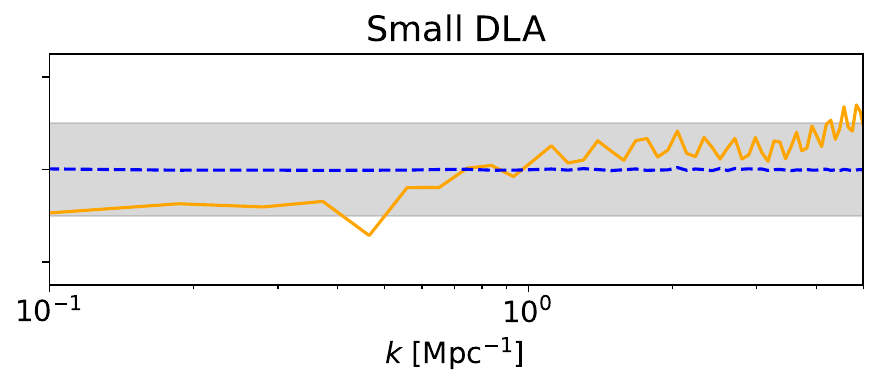}
    \caption{Top: \poned measured from the double-line masked spectra (green points) compared to two models. The orange line shows the measurement from unmasked spectra, which we consider the `original' model, and the blue dashed line shows the model after applying the window matrix to the original model, where the window matrix incorporates the double-line mask. The lower panel of the top figure shows the percentage difference between measurement and model when comparing with the original model (orange) and masked model (blue). The residuals are well within the 1\% shaded regions when the mask is modeled, i.e., applying the window function accurately replicates the masking impacts. Bottom: the percentage residuals are displayed for the four other masking cases.}
    \label{fig:model-mask-double-line}
\end{figure}

\Cref{fig:model-mask-double-line} reproduces the mask impacts and also shows the performance of the window-convolved model for each case. Here, we calculate the window matrix from \cref{eq:M_mn} using each set of masks and apply it directly to the unmasked \poned measurement $P_\mathrm{1D,orig}$. We choose to use $P_\mathrm{1D,orig}$ as the `theory model', rather than a smoothed theory model, in order to remove the effects from cosmic variance in the comparison between model and measurement. Comparing the masked measurements with the original model reveals $>1\%$ biases within $k<5~\iMpc$ for all masks with small-scale features (orange lines). However, the model including the window function reproduces the masked signal to within $0.5\%$ in a scale-independent manner (blue dashed lines). Building from the tests with Gaussian skewers in \cref{ss:nonperiodic}, these validation tests with hydrodynamic mock data further demonstrate that comparison of a masked measurement with a masked model can yield robust, unbiased inference for a wide range of masking types, even when the original power spectrum has features.

\section{Masking in FFT estimators of \pcross} \label{sec:px}
\subsection{Overview}
While FFT estimators have been widely used in measurements of $\poned(k_\parallel)$, the geometry of the \lya forest surveys makes it difficult to apply them to measurements of the 3D power spectrum, $\pthreed(k_\perp, k_\parallel)$. 
Each quasar spectrum gives us information about the \lya forest fluctuations along the line of sight, with a fairly homogeneous window matrix (except for the occasional masked pixels discussed above). 
However, the window function is very complicated in the transverse direction, where we only sample the 3D field whenever we happen to have a background quasar.

This has motivated the use of alternative methods to measure 3D correlations \cite{FontRibera2018,AbdulKarim2024,deBelsunce2024,Horowitz2025, Karacayli2025}.
An interesting idea, first proposed in \cite{Hui1999}, is to measure the cross-spectrum, $\pcross(r_\perp,k_\parallel)$, the correlation of 1D Fourier modes $\delta(k_\parallel)$ in neighboring quasars separated by a transverse separation $r_\perp$.
This summary statistic captures the same information contained in \pthreed, while not being affected by the complex transverse sampling. 
In the continuous, static (no redshift evolution) limit, it is defined as:
\begin{equation} \label{eq:px_def}
    \left< \delta(x_\perp, k_\parallel) ~ \delta^\ast(x^\prime_\perp, k_\parallel^\prime) \right> = 2 \pi ~ \delta^D(k_\parallel-k_\parallel^\prime) ~ \pcross(r_\perp, k_\parallel) ~, 
\end{equation}
with $r_\perp = | x^\prime_\perp - x_\perp |$. 
It is easy to see that in the limit of $r_\perp \rightarrow 0$, $\pcross(r_\perp=0,k_\parallel)$ is equivalent to $\poned(k_\parallel)$. 
\pcross is related to \pthreed through the inverse Fourier transform of the perpendicular modes $\boldsymbol{k}_\perp=(k_{\perp,x},k_{\perp,y})$ for a particular $r_\perp$ transverse scale:
\begin{equation}\label{eq:px_to_3D}
    P_{\times}(k_{\parallel}, r_\perp) = 
    \frac{1}{2\pi} \int_0^{\infty}  \mathrm{d}k_\perp\,  k_\perp\, J_0 (k_\perp r_\perp) \, P_\mathrm{3D}(k_\parallel, k_\perp),
\end{equation}
where $J_0$ is the Bessel function of the first kind. At zero separation, $J_0(r_\perp=0)=1$, giving the relation between \pthreed and \poned. We note that \cref{eq:px_to_3D} neglects to consider redshift evolution
(see \cite{FontRibera2018} for a discussion on how to define \pcross in an evolving universe). 

The optimal quadratic estimator to measure \pcross, discussed in \cite{FontRibera2018}, is quite complex and computationally expensive.
On the other hand, the FFT estimators used to measure \poned can be easily extended to measure \pcross.
Recently, \cite{AbdulKarim2024} used an FFT estimator to measure \pcross from the final \lya dataset of eBOSS.
In this section we show that the methodology described in \cref{sec:p1d_math} can be trivially extended to \pcross, enabling the use of weights and the possibility to forward model the impact of masked pixels.

\subsection{\pcross from pairs of quasars with different redshifts}

\begin{figure}
    \centering
    \includegraphics[width=0.95\linewidth]{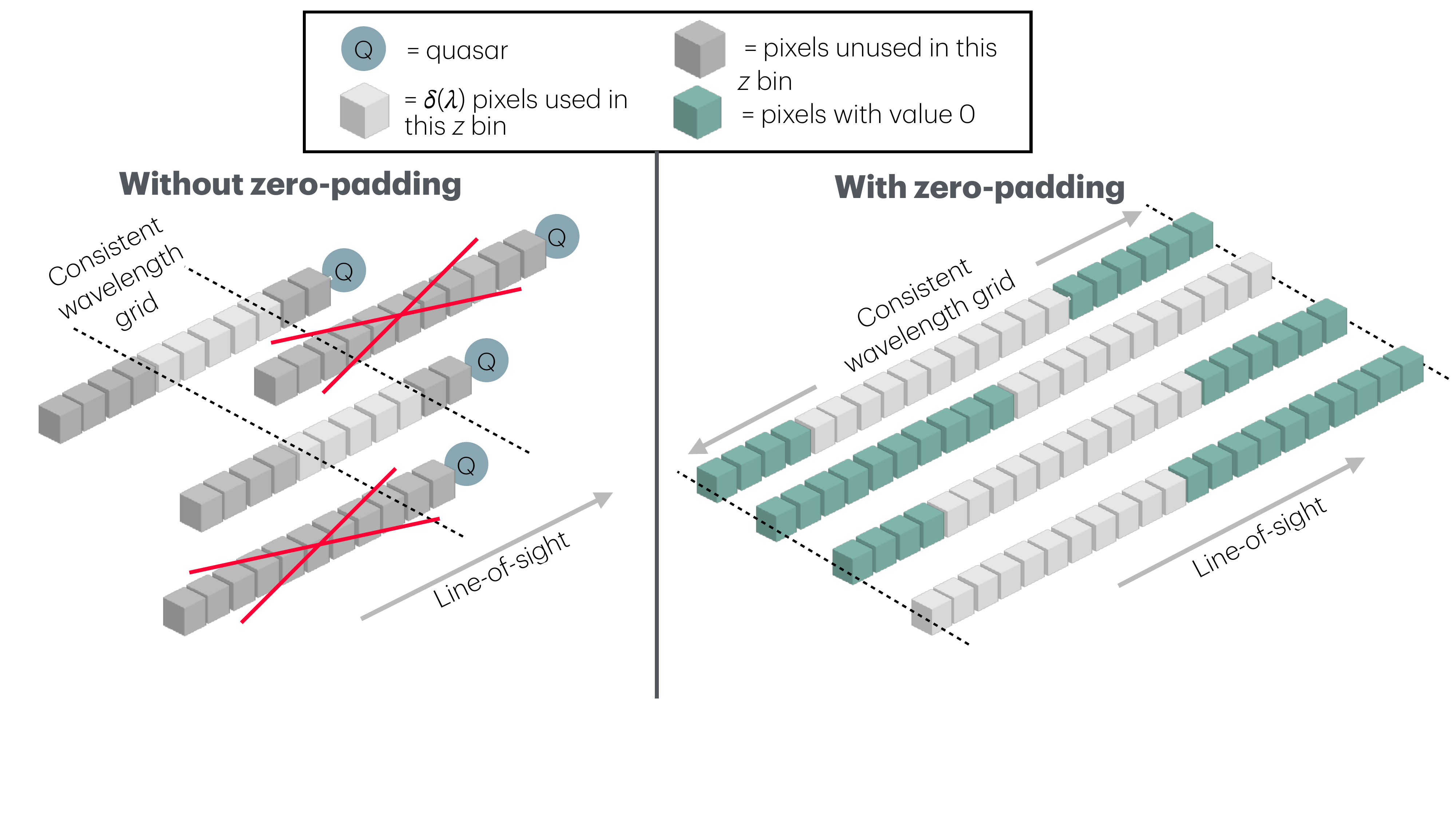}
    \caption{A visualization of the way in which zero-padding can be used to preserve data when measuring \pcross. At left, a limited wavelength (redshift) grid is defined, and two of the four spectra are removed because they do not fully span the grid. The shaded pixels beyond the grid are also unused in the measurement. At right, a larger wavelength (redshift) range is defined to incorporate more data. The addition of zeros places all spectra into a consistent wavelength grid, corresponding to a consistent FFT grid once Fourier transformed. All pixels are used, so the impacts of zeros must be modeled at the measurement or inference stage.}
    \label{fig:grid-visuals}
\end{figure}

The discussion of zero-padding from Section ~\ref{sec:p1d_math} is particularly relevant to \pcross, as the measurement of correlation between skewer pairs via the FFT estimator requires that both skewers in each pair share the same FFT grid. Observationally, this is challenging, because the \lya skewers are not necessarily the same length, and the redshifts of the quasars (marking the start of each skewer) vary. 
\cite{AbdulKarim2024} tackled this issue by defining a limited common wavelength range for the \pcross measurement, and removing data that falls beyond that range and spectra that only partially covered it. 
Rather than removing this information and thus depleting the statistical power of the measurement, it is possible to retain all the data by adding zero-padding to each spectrum such that all fall along a common grid in wavelength space, which is then converted into a common grid in Fourier space. In \cref{fig:grid-visuals} we show a visual representation of this change in methodology.
As discussed in \cref{ss:nonperiodic}, the zero-padding is also necessary to efficiently model the impact of masking, pixel weights, and non-periodic skewers. 

Using the method described in \cref{sec:p1d_math}, we generate $10^7$ pairs of periodic, Gaussian \lya skewers with Fourier modes ($\delta^A(k), \delta^B(k)$, where $k$ refers to one-dimensional modes, called $k_\parallel$ in the previous sub-section).
In order for them to have a non-zero \pcross, we generate a new set of Fourier modes $\delta^C(k)$ as a linear combination of these two, $\delta^C = f_\times \delta^A + \sqrt{1-f_\times^2} \delta^B$.
Computed this way, \poned of fields $\delta_A$ and $\delta_C$ is equal to the input power detailed in \cref{tab:true_p1d}, and the cross-spectrum of $\delta^A$ and $\delta^C$
will be equal to $\pcross = f_\times \poned$. We set  $f_\times = 0.7$.

Following \cref{sec:p1d_math}, in order to break the periodicity of the skewers we only keep 256 pixels ($1/8$ of the original skewer), and add a factor of two zero-padding to obtain an array of 512 pixels, half of them masked ($w_a=0$). 
In order to simulate the fact that each line-of-sight covers a different range of observed wavelengths (set by the redshift of the background quasars), each simulated skewer has a different starting point\footnote{To be precise, we draw a random number between 0 and 256 to be used as a starting point for the 256 pixels that are kept.}.

For each pair, we then measure the correlation of 1D Fourier modes from the two skewers $I$ and $J$, with masks $w^I_a$ and $w^J_a$, respectively, and separated by $\theta_{IJ}$ 
\footnote{It is important to note that the two skewers share the same FFT grid, with the quasar redshift determining where the padded zeros should be placed.}. 
We will refer to their discrete, masked Fourier modes as $f^I_m$ and $f^J_m$, respectively, defined as in \cref{eq:f_m}.
Analogously to \cref{eq:fourier_mode_variance}, the correlation of these modes is:
\begin{equation}
 \left< f^I_m ~ f_m^{J\ast} \right>      
    = \frac{\Delta k}{2\pi} \sum_n w^I_{m-n} ~ w_{m-n}^{J\ast} ~ R^2(k_n) \pcross(\theta_{IJ}, k_n) ~,
\end{equation}
and analogously to \cref{eq:masked_fft} we can write an estimator for \pcross (in some narrow $\theta$ bin $A$) as:
\begin{align}
 \hat P^A_{\times,m} 
 & \equiv C^A_m \frac{\Delta x}{N ~ R^2(k_m)} \frac{1}{N_A} \sum_{\substack{I,\, J \in A \\ I < J}}
 \left( f^I_m ~ f_m^{J\ast} + f^{I\ast}_m ~ f_m^J \right)   \nonumber \\ 
 & = C^A_m \frac{\Delta x}{N ~ R^2(k_m)} \frac{2}{N_A} \sum_{\substack{I,\, J \in A \\ I < J}}
 \Re \left( f^I_m ~ f_m^{J\ast} \right)  ~,
\end{align}
where the sum runs over all pairs of skewers within the transverse separation bin $A$ without double-counting.

The ensemble average of the estimator is related to the original \pcross by:
\begin{align}
 \left< \hat P^A_{\times,m} \right>
    & = C^A_m ~ \frac{\Delta x}{N ~ R^2(k_m)} \frac{2}{N_A} 
        \sum_{\substack{I,\, J \in A \\ I < J}}
 \left< \Re \left( f^I_m ~ f^{J\ast}_m \right) \right> \nonumber \\
    & = \frac{C^A_m}{N^2 ~ R^2(k_m)} \frac{2}{N_A} \sum_{\substack{I,\, J \in A \\ I < J}}
        \sum_n \Re \left( w^I_{m-n} ~ w^{J\ast}_{m-n} \right) 
        R^2(k_n) ~ \pcross(\theta_A, k_n)   \nonumber \\
    & = \frac{C^A_m}{N^2 ~ R^2(k_m)} \sum_n W^A_{m-n} ~  
        R^2(k_n) ~ \pcross(\theta_A, k_n)   \nonumber \\
    & = \sum_n M^A_{mn} ~ \pcross(\theta_A, k_n) ~,
\end{align}
where we have introduced $W^A_m$ as:
\begin{equation}
 W^A_m \equiv \frac{2}{N_A} \sum_{\substack{I,\, J \in A \\ I < J}}
 \Re \left( w^I_m ~ w^{J\ast}_m \right) 
\end{equation}
and we have introduced an equivalent window matrix for \pcross:
\begin{equation}
 M^A_{mn} \equiv \frac{C^A_m}{N^2 ~ R^2(k_m)} W^A_{m-n} R^2(k_n) ~.
\end{equation}

Here again we can choose the normalization factor $C_m^A$ such that $1 = \sum_o M^A_{mo}$. 
This condition is satisfied if we choose:
\begin{equation}
 C^A_m = \frac{N^2 R^2(k_m)}{\sum_o W^A_{m-o} R^2(k_o)} ~.
\end{equation}

Using this normalization factor, the estimator can be rewritten as:
\begin{equation}
 \hat P^A_{\times,m} \equiv \frac{L}{\sum_o W^A_{m-o} R^2(k_o)} ~ \frac{2}{N_A} \sum_{\substack{I,\, J \in A \\ I < J}}
  \Re \left( f^I_m ~ f^{J\ast}_m \right) ~,
\end{equation}
and the window matrix as:
\begin{equation}\label{eq:window_matrix_px}
 M^A_{mn} \equiv \frac{W^A_{m-n} R^2(k_n)}{\sum_o W^A_{m-o} R^2(k_o)}  ~.
\end{equation}

\begin{figure}
\centering
 \includegraphics[width=0.8\textwidth]{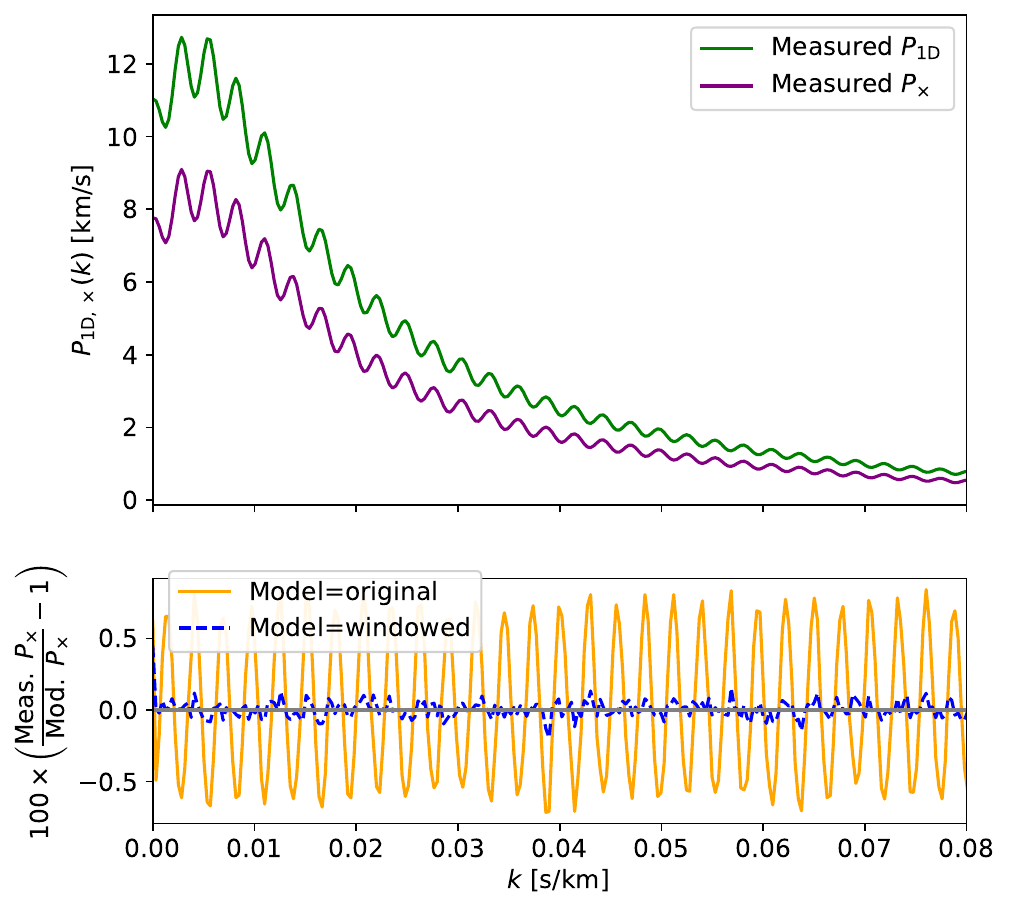}
 \caption{(Top): measured \poned and \pcross from $10^7$ pairs of Gaussian skewers.
 (Bottom): residuals of the measured \pcross from the same pairs of \lya skewers compared to the true \pcross used to generate the mocks (orange), compared to the prediction that uses the window matrix in \cref{eq:window_matrix_px} to forward-model the impact of masking and zero padding (blue, dashed). 
 }
 \label{fig:convolved_px}
\end{figure}

The measurement of \poned and \pcross from these pairs of Gaussian skewers is shown in the upper panel of \cref{fig:convolved_px}. The lower panel compares the \pcross measurement to the original model and also to the windowed model. Similarly to the \poned demonstrations of previous sections, application of the window matrix faithfully reproduces the distortions that zero-padding causes to the original power spectrum, reducing the residuals by a factor of $\sim10$. The exercise demonstrates that in cases of different skewer starting locations, it is possible to retain all the data to optimize signal-to-noise for a \pcross measurement, by incorporating zero-padding and modeling it in predictions.

\section{Conclusion} \label{sec:conclusion}

This work presented an efficient method to estimate and robustly model the \poned and \pcross statistics of the \lyaf with FFTs in the presence of spectral pixel masking and pixel weights. 
We first demonstrated how the non-periodicity and finite length of realistic spectra induce mode-mixing in the power spectra, which causes scale-dependent bias and smooths features.
We described an estimator which zero-pads the spectra by at least a factor of two in order to make the $k$ grid finer and enable the distortions to be modeled rapidly using FFTs. 
We then translated the calculations of a window matrix given masking---already commonly used in galaxy surveys, the CMB, and other \lya contexts to describe sky incompleteness and spectral pixel masking---to the specific domain of FFT estimators for \lya \poned and \pcross. 
Using Gaussian and hydrodynamic simulations, we demonstrated that the model accounts for biases from zero-padding and other varied types of realistic masking at high accuracy. 
This combined estimator and modeling framework enables precision inference with DESI data and other future surveys.

In practice, the window matrix calculation could be applied in an inference pipeline through two different approaches. 
In one approach, the theory prediction gets contaminated with realistic models for contaminants, convolved with the same window matrix that affects the data, and then compared to observational measurements. (Subsequently, parameters are iterated through with a sampling technique such as MCMC.) 
Alternatively, in some settings it might be possible to invert the window matrix in order to directly deconvolve it from the measurement.
In that case, one would not need to forward-model the masking or resolution effects on the theory prediction, but rather compare the original prediction directly to the measurement. An additional advantage to this approach is the ability to compare between an optimal quadratic estimator and an FFT estimator. On the other hand, for either 
estimator, inverting the window requires making 
unnecessary assumptions
about the discretization of the theory, and generally 
leads to 
anti-correlated error covariance. E.g., past OQE
work advocated an intermediate transformation of the window
which left the errors uncorrelated \citep{sdss2004ApJ...606..702T}.
In any case, given a well-computed window matrix
we can easily explore different ways to use it. 

An important feature of the weighted FFT estimators discussed in this work is the ability to optimize them using inverse-variance pixel weights. With an accurate model for the window matrix in hand, one can down-weight high-variance pixels to place more importance on well-measured pixels, and model the resulting impacts on \poned or \pcross by calculating the window matrix with that array of weights. 
Pixel weights have thus far not been used in the \pfft estimators, and their inclusion will help maximize the information gained from DESI \lya spectra.


There are several limitations to the methods recommended in this work. 
First, this paper does not address the complication of masks that correlate with the signal. Correlated masks will complicate the measurement and modeling of any estimator (not only FFT-based). This is true of DLA masking, where the removal of dense absorbers also removes \lyaf signal at biased locations. When masking only the largest DLAs, highly extended masks are typically used, removing broad regions of the forest that include a wide range of densities and are thus nearly unbiased on average. 
However (somewhat counterintuitively) masking small DLAs causes more bias to the signal, as the masks are shorter and therefore remove a shorter and less representative region. Analyses masking only large DLAs can likely ignore the correlation and use the formalism presented here, but must model the bias to the power spectra given by the inclusion of residual small DLAs. 
Those masking a broader range of DLA sizes take care of the latter bias, but then face the problem of a more complicated masking bias which cannot use the approach of this paper. 
Such cases likely need to be handled by modeling via hydro simulations, which is beyond the scope of this paper.

A second limitation is that we have elected not to model a full resolution matrix in the development of the model, rather assuming that each pixel has equal resolution which is a symmetric function. Further, we have used perfect resolution ($R=1$) in the tests with Gaussian and hydro skewers, electing to use the native simulations rather than smoothing to a realistic resolution and accounting for it in the models. In any real observational data, resolution varies by pixel and by spectrum and the resolution functions generally may be asymmetric. Depending on the size of these impacts, it may be necessary to incorporate the full resolution matrix in models for \poned or \pcross. This is beyond the scope of this work.

This work presents a key step toward precision modeling of small-scale \lyaf statistics. With a full pipeline incorporating accurate pure-theory predictions, inclusion of metals and other contaminants, and propagation of the mask, weights, and resolution effects through the estimator, future analyses will be able to maximize the information gained from the vast amounts of incoming spectroscopic data to infer high-redshift cosmology at unprecedented precision.

\section*{Acknowledgements}

The research of ML and AFR was supported by the European
Union’s Horizon Europe research and innovation programme
(COSMO-LYA, grant agreement 101044612) and by the Spanish Ministry of Science and Innovation (project PGC2021-123012NB-C41). 
AFR acknowledges funding by the Spanish Ministry of Science and Innovation under the Ramon y Cajal program (RYC-2018-025210).
IFAE is partially funded by the CERCA program of the Generalitat de Catalunya. PM is supported by the U.S. 
Department of Energy (DOE), Office of Science, Office of
High-Energy Physics, under Contract No. DE-AC02-05CH11231.

\bibliographystyle{JHEP.bst}
\bibliography{main}

\newpage
\appendix

\end{document}